\documentclass[letterpaper,journal]{IEEEtran}
\usepackage{amsmath,amsfonts}
\usepackage{algorithmic}
\usepackage{algorithm}
\usepackage{array}
\usepackage[caption=false,font=normalsize,labelfont=sf,textfont=sf]{subfig}
\usepackage{textcomp}
\usepackage{stfloats}
\usepackage{url}
\usepackage{verbatim}
\usepackage{graphicx}
\usepackage{cite}
\usepackage{xcolor}
\usepackage{tabularx}
\usepackage{booktabs}
\usepackage{tcolorbox}

\usepackage{orcidlink}
\hypersetup{hidelinks}
\newcommand{\orcid}[1]{\,\orcidlink{#1}} 
\begin{document}

%\title{A Sample Article Using IEEEtran.cls\\ for IEEE Journals and Transactions}
\title{TensorMutation: Learning Weight-Space Mutations of Deep Learning Models on Neural Graphs}

\author{Zaheed Ahmed\orcid{0000-0001-6594-537X}, Philip Makedonski\orcid{0000-0001-7752-0029}, and Jens Grabowski\orcid{0000-0003-2994-3531}
        % <-this % stops a space

%\thanks{Manuscript received April 19, 2021; revised August 16, 2021.}
\thanks{Zaheed Ahmed, Philip Makedonski, and Jens Grabowski are with the Institute of Computer Science, University of Göttingen, Germany (e-mail: zaheed.ahmed@informatik.uni-goettingen.de; makedonski@informatik.uni-goettingen.de; grabowski@informatik.uni-goettingen.de).}}

\markboth{Journal of \LaTeX\ Class Files,~Vol.~14, No.~8, August~2021}%
{Shell \MakeLowercase{\textit{et al.}}: A Sample Article Using IEEEtran.cls for IEEE Journals}


\maketitle

\begin{abstract}
Mutation testing for deep learning (DL) faces a cost--realism trade-off.
The pre-training approach produces realistic but computationally expensive mutants, whereas existing post-training operators modify trained models efficiently but produce mutants that less closely resemble real faults.
This study asks whether the trade-off is necessary: can the weight-space effect of a fault-informed pre-training mutation operator be predicted from a trained model, without retraining it?
We answer this question with TensorMutation, which represents a trained model as a neural graph and uses a mutation-conditioned graph neural network (GNN) to predict parameter changes.
The GNN is trained offline and reused across unseen models so that TensorMutation can generate executable mutants without retraining.

We construct a corpus of 25\,252 original--mutant pairs from 71 multilayer perceptron (MLP) classifiers using 11 fault-informed pre-training operators, including 25 newly curated subjects.
TensorMutation partially reproduces the behavioral effects of pre-training mutations on nine held-out subjects, and direct weight-space supervision significantly improves weight-space fidelity.
On 12 real-fault subjects, TensorMutation achieves the highest median behavioral similarity among the evaluated approaches.
The improvement over the post-training tool DeepMutation++ is statistically significant (Holm-adjusted $p=0.018$), whereas the differences from the pre-training tool DeepCrime and the norm-matched random control are not significant.
These results show that fault-informed pre-training mutations leave learnable structure in weight space that transfers to unseen models.
\end{abstract}

\begin{IEEEkeywords}
Post-training mutation, weight-space learning, mutation testing, deep neural networks, neural graphs, graph neural networks.
\end{IEEEkeywords}

\section{Introduction}
\label{sec:introduction}

\IEEEPARstart{M}{utants} are commonly used as proxies for real faults to evaluate the effectiveness of test suites~\cite{just_are_2014}.
In deep learning (DL), their use extends beyond serving as test targets.
Mutants have been employed to evaluate automated program repair~\cite{wu_mutation-based_2021, sohn_arachne_2023}, robustness and adversarial testing~\cite{hu_deepmutation_2019, lin_robustness_2022, mendez_testing_2024, wang_adversarial_2019, pour_search-based_2021}, test generation, improvement, and prioritization~\cite{wang_prioritizing_2021, riccio_deepmetis_2021, deokuliar_improving_2024}, data cleaning~\cite{li_cleanml_2021, abdelaal_rein_2023}, fault localization~\cite{cao_deepfd_2022, ghanbari_mutation-based_2023, raza_graph_2025}, evaluation of metamorphic testing techniques~\cite{xie_testing_2011}, and several other testing and debugging tasks~\cite{wang_deep_2020, ghanbari_decomposition_2024, jahangirova_quality_2021}.
Across these applications, the usefulness of mutation-based evaluation depends on mutants being representative of faults that occur in real DL systems.
At the same time, mutant generation must remain computationally practical if mutation testing is to be applied at scale.

In DL, mutation approaches are commonly divided into \emph{pre-training} and \emph{post-training} mutation according to the stage at which the mutation operator is applied~\cite{ma_deepmutation_2018, ahmed_exploring_2024}.
Pre-training mutation injects faults into the training program or training data and then retrains the model to obtain a mutant~\cite{ma_deepmutation_2018, humbatova_deepcrime_2021}.
Post-training mutation instead applies mutation operators directly to an already trained model, typically by perturbing its parameters or architecture~\cite{ma_deepmutation_2018, shen_munn_2018, hu_deepmutation_2019, kim_muff_2025}.
The cost of pre-training mutation becomes particularly high when multiple independently trained instances of each mutant are required to account for the stochastic nature of DL training, as in the statistical killing criterion~\cite{jahangirova_empirical_2020} used by DeepCrime~\cite{humbatova_deepcrime_2021, humbatova_deepcrime_2023}.

In a preliminary controlled comparison on LeNet-5/MNIST, Abbasishahkoo et al.~\cite{abbasishahkoo_teasma_2024} reported that mutation analysis required less than 49 minutes with DeepMutation++~\cite{hu_deepmutation_2019}, compared with more than 123 hours with DeepCrime, a difference of more than 150 times.
Although this comparison involved only one modest-sized subject, it illustrates the substantial computational overhead introduced by retraining-based mutation.
At the same time, our previous empirical study~\cite{ahmed_empirical_2025} found DeepCrime mutants to resemble real faults more closely than DeepMutation++ mutants.
Established post-training approaches avoid retraining through manually defined transformations of trained models rather than transformations learned from fault-informed pre-training mutations.
Together, these findings motivate an approach that approximates the effects of fault-informed pre-training mutations while generating mutants without retraining.

This study asks whether pre-training mutations leave learnable structure in the parameters of trained models.
A pre-training mutation changes the training process and can therefore alter the learned parameters of the resulting model.
Shah et al.~\cite{shah_towards_2025} found systematic differences in the weights and biases of models trained with data faults, providing evidence that data faults can leave detectable parameter-space signatures.
We therefore investigate whether the parameter changes induced by fault-informed pre-training mutation operators can be predicted for previously unseen models.
The idea follows from our earlier finding for conventional programs that even chains of manually defined mutations fail to reproduce most real faults, which led us to propose learning mutations from real faults instead~\cite{ahmed_new_2024}.
We call this approach \emph{TensorMutation} because its learned transformation \emph{mutates} the trained model's weight and bias \emph{tensors}.

TensorMutation represents a trained model as a neural graph~\cite{kofinas_graph_2024} and conditions a graph neural network (GNN) on the requested mutation specification through Feature-wise Linear Modulation (FiLM)~\cite{perez_film_2018} to predict parameter changes.
It learns this transformation from pairs of original models and their pre-training mutants generated by the fault-informed operators of DeepCrime~\cite{humbatova_deepcrime_2021}.
To generate a mutant without retraining, TensorMutation applies the predicted changes directly to the trained model and performs any structural modifications required by the specified mutation.
Thus, mutant generation is post-training, while the learned transformation is grounded in the effects of pre-training mutation.
The GNN is trained beforehand on a corpus of such pairs, so mutant generation for unseen models requires only GNN inference and mutant reconstruction.

Ideally, TensorMutation would learn directly from reproducible real bug-fix pairs.
However, obtaining such pairs at the scale required for training remains difficult in DL.
Jahangirova et al.~\cite{jahangirova_real_2026} were able to reproduce only 54\% of the faults considered from existing DL fault datasets, and Morovati et al.~\cite{morovati_bugs_2023} found that only about 3\% of the machine-learning-related bugs they investigated from GitHub and Stack Overflow were reproducible.
Moreover, real faults lack a shared vocabulary: there is no agreed way for a user to request a particular fault type when generating a mutant.
Fault-informed mutation operators provide a practical alternative.
Each operator represents a fault type documented in empirical studies and provides an explicit specification for generating that type of fault.
The same operator can also be applied across subjects and configurations to produce many training pairs.
TensorMutation therefore learns from mutants generated by fault-informed pre-training operators rather than directly from real faults.
These mutants form synthetic bug-fix pairs, where the original model serves as the fixed version and the mutant as the faulty version, while preserving the fault type represented by the operator.
We collect these pairs at scale into the corpus released with this work.

In this study, a subject is a TensorFlow/Keras program that trains a multilayer perceptron (MLP) classifier on its corresponding dataset.
We evaluate TensorMutation across 71 subjects and 11 DeepCrime operators, with nine subjects held out for testing, and on 12 real-fault subjects.
The evaluation asks four research questions (RQs) that build on one another.
RQ1 and RQ2 examine how faithfully TensorMutation reproduces the corresponding pre-training mutants on unseen subjects, in behavior and in weight space.
RQ3 then asks whether behavioral fidelity and weight-space fidelity are related.
Finally, RQ4 investigates how closely TensorMutation mutants resemble real faults compared with mutants generated by DeepCrime, DeepMutation++, and norm-matched random perturbations, which preserve the magnitude of the predicted change but randomize its direction.
We additionally examine, as a design ablation, whether conditioning the GNN on the source model's training configuration improves the learned transformation.
The main contributions of this work are:

\begin{itemize}

\item We introduce TensorMutation, a learned post-training mutation approach that predicts the weight-space effects of fault-informed pre-training mutations and generates executable mutants without retraining.

\item We provide empirical evidence that mutation-induced parameter changes contain learnable structure that transfers to unseen subjects and that direct weight-space supervision improves weight-space fidelity.
On 12 real-fault subjects, TensorMutation achieves significantly higher behavioral similarity to real faults than DeepMutation++.

\item We release a corpus of 25\,252 original--mutant pairs from 71 subjects and 11 fault-informed pre-training operators, including 25 newly curated MLP subjects, to support research on testing and debugging of DL systems.

\end{itemize}

The remainder of the paper is organized as follows.
Section~\ref{sec:related-work} reviews DL mutation testing and weight-space learning literature in relation to this work.
The TensorMutation approach is presented in Section~\ref{sec:approach}.
Section~\ref{sec:study-design} describes the study design and its four RQs.
The results for the RQs, followed by the source-context design ablation, are reported in Section~\ref{sec:results}.
Section~\ref{sec:discussion} discusses the findings and their implications, and Section~\ref{sec:threats-to-validity} examines threats to validity.
Finally, Section~\ref{sec:conclusion} concludes the study and outlines future directions.

\section{Related Work}
\label{sec:related-work}

\noindent This study lies at the intersection of two independent fields, mutation testing of DL systems and weight-space learning.
We review each in turn and then position our work.

\subsection{Mutation Testing of DL Systems}
\label{sec:mutation-testing-dl-systems}

The literature on mutation testing of DL systems relevant to our work can be grouped along two axes: how mutants are generated and how realistic they are, and how the cost of mutation testing can be reduced.

\subsubsection{Mutation Approaches}
Mutation analysis was carried over to neural networks in 2018 by MuNN~\cite{shen_munn_2018} and DeepMutation~\cite{ma_deepmutation_2018}.
DeepMutation distinguished mutations applied before training from those applied directly to trained models, where a mutant costs a perturbation rather than a training run.
DeepMutation++~\cite{hu_deepmutation_2019} extended the post-training operators to recurrent networks.
A systematic empirical evaluation of these operators~\cite{jahangirova_empirical_2020} found that their effectiveness depends strongly on configuration, with some configurations yielding mutants that are trivial or impossible to kill.
Recently, MuFF~\cite{kim_muff_2025} introduced manually defined weight- and neuron-inhibitor operators together with an automated stability check to obtain mutants that are stable across runs and sensitive to test-set quality.
Unlike these approaches, DeepCrime~\cite{humbatova_deepcrime_2021, humbatova_deepcrime_2023} grounds its operators in empirical studies of real faults~\cite{zhang_empirical_2018, islam_comprehensive_2019, humbatova_taxonomy_2020}.
Because those faults occur in the training program or data, DeepCrime is a pre-training technique: every mutant is retrained, and killing is decided statistically over repeated trainings to absorb nondeterminism.

The assumption that mutants stand in for real faults was carried into DL without experimental verification.
Our empirical comparison~\cite{ahmed_empirical_2025} introduced a statistical framework for coupling strength and behavioral similarity to real faults and found that DeepCrime mutants showed significantly stronger coupling and higher behavioral similarity than DeepMutation++ mutants.
Realism therefore lies on the retraining side of this trade-off, the side that supplies the corpus TensorMutation learns from.

\subsubsection{Cost Reduction Techniques}
A separate line of work accelerates DL mutation testing without changing how mutants are generated.
Some techniques execute fewer test inputs, for example by sampling inputs near the decision boundary~\cite{shen_boundary_2021}.
Others generate or execute fewer mutants.
Operator groups can be reduced based on the decision-boundary changes induced by their mutants~\cite{feng_mutation_2022}, while quality-based criteria have been used to prioritize mutation operators~\cite{wang_fine-grained_2023, zhang_fine-grained_2025} and operator configurations~\cite{ahmed_quality-driven_2026}.
Strongly subsuming second-order mutants can further replace multiple first-order mutants~\cite{li_how_2022}.
Clustering neurons or mutants with similar weights also reduces the number of mutants generated or tested~\cite{lyons_accelerating_2025}.
Xue et al. make each execution cheaper by detecting behavioral deviations earlier through weak mutation~\cite{xue_deepweak_2024}.
All of these reduce how many inputs or mutants are executed, or how much each execution costs; none reduces the cost of producing a pre-training mutant, which is the cost of retraining.
TensorMutation instead targets the cost of producing each mutant by replacing retraining with GNN inference and is complementary to these techniques.

\subsection{Weight-Space Learning}
\label{sec:weight-space-learning}

Treating trained neural networks as a new data modality has emerged as a field of its own, organized into understanding the symmetries of parameter space, representing models as embeddings, and generating or transforming weights.
We refer readers to Han et al.~\cite{han_survey_2026} for a general overview and confine ourselves here to the two strands TensorMutation builds on.

\subsubsection{Weight-Space Representation}
The idea behind weight-space representation is that properties of a trained network can be inferred from its parameters.
Network accuracy can be predicted from weight statistics without evaluating the network on input data~\cite{unterthiner_predicting_2021}, properties of the training setup can be classified from weights~\cite{eilertsen_classifying_2020}, and self-supervised hyper-representations learned over model zoos capture both~\cite{schurholt_self-supervised_2021}.
These findings suggest, but do not establish, that faults introduced during training may also leave learnable signatures in the resulting weights, which is the possibility TensorMutation investigates.
Weight-space representations need to consider symmetry, since networks that differ only by a consistent permutation of hidden neurons compute the same function~\cite{entezari_role_2021, ainsworth_git_2023}.
Early equivariant architectures explicitly encoded permutation symmetries for specific network families~\cite{navon_equivariant_2023, zhou_permutation_2023}.
Graph-based approaches represent trained neural networks as graphs and process them with permutation-equivariant GNNs~\cite{kofinas_graph_2024, lim_graph_2024}.
TensorMutation adopts this graph representation.

\subsubsection{Weight-Space Editing}

Weight editing changes the parameters of an existing model and can be expressed as $\tilde{\theta} = \theta + \Delta(\theta)$, where $\Delta$ is a mapping over the weights~\cite{han_survey_2026}.
Existing post-training operators that perturb parameters are hand-designed instances of such a mapping.
In the weight-space learning literature, $\Delta$ is instead learned from data, and TensorMutation adopts this formulation.
Permutation-equivariant metanetworks have been used to edit implicit neural representations directly in weight space~\cite{kofinas_graph_2024, lim_graph_2024, zhou_permutation_2023}, and interpretable edit directions have been identified in the weight space of customized diffusion models~\cite{dravid_interpreting_2024}.

Two findings from weight generation and reconstruction are relevant to learning such edits.
Meynent et al.~\cite{meynent_structure_2025} show that combining structural and behavioral supervision improves neural network weight reconstruction.
TensorMutation investigates a related principle for fault-induced transformations: its joint objective combines parameter-delta and behavioral supervision to approximate a specified mutation, rather than to preserve the original model's behavior.
Some generative weight models largely reproduce or interpolate their training checkpoints rather than generalize beyond them~\cite{zeng_generative_2026}, which motivates our evaluation on subjects held out from training.
A closely related direction is weight forecasting: NiNo predicts future weights from recent parameter states using a GNN over neuron connectivity~\cite{knyazev_accelerating_2025}.
This shows that training-related parameter updates can be learned, but does not establish that changes caused by training faults are similarly predictable, which is the question TensorMutation investigates.

To our knowledge, no prior work in weight-space learning learns to change a trained model the way a specified fault would, and no DL mutation approach learns its operators as weight-space transformations.
TensorMutation connects the two fields: it learns the weight-space effect of a fault-informed pre-training mutation and applies it to a trained model to generate mutants without retraining.

\section{TensorMutation Approach}
\label{sec:approach}
\noindent This section first defines the learning problem and provides an overview of TensorMutation.
It then describes the main components of the approach.

\subsection{Problem Formulation and Overview}
\label{sec:problem-formulation-and-overview}

Let $M_o=(A_o,\theta_o)$ denote a trained original model, where $A_o$ is its architecture and $\theta_o$ its trainable parameters.
A mutation specification $m=(\textit{operator},\textit{configuration})$ identifies both the pre-training mutation and the concrete setting in which it is applied.
For example, the operator may change the learning rate, while the configuration gives the target value.
A pre-training mutation tool realizes $m$ by retraining the model under the mutated configuration.
This produces a mutant $M_m=(A_m,\theta_m)$ and the corresponding parameter change $\Delta\theta_m=\theta_m-\theta_o$.
TensorMutation learns to approximate this parameter change directly from the trained model and the requested mutation.
It uses a GNN to process the model as a neural graph because message passing follows the model's neuron--connection structure and can operate on subjects with different numbers of nodes and edges.
Formally, it learns the conditional transformation
\begin{equation}
f_\phi:\left(G(M_o,m),m\right)\longmapsto\widehat{\Delta\theta}_m,
\end{equation}
where $G(M_o,m)$ is the mutation-aware neural graph, $\phi$ denotes the trainable parameters of the GNN, and $\widehat{\Delta\theta}_m$ is the predicted parameter change.
The specification $m$ appears twice because it provides local mutation information in the graph $G(M_o,m)$ where applicable (Section~\ref{sec:neural-graph}) and conditions the GNN on which transformation to predict (Section~\ref{sec:mutation-conditioning}).
The transformation is learned from a corpus of $(M_o,m,M_m)$ training tuples as described in Sections~\ref{sec:training-corpus-construction}--\ref{sec:training-gnn}.
The mutant $M_m$ provides two kinds of supervision: its behavior on fixed probe inputs, which is the learning target, and its parameter change $\Delta\theta_m$, which can additionally supervise the prediction (Section~\ref{sec:learning-objective}).
We train the GNN beforehand, and at generation time a user supplies $M_o$ and $m$.
TensorMutation then applies the predicted parameter change $\widehat{\Delta\theta}_m$ to $\theta_o$ together with any deterministic structural modification required by $m$.
The mutant is therefore produced without retraining.

\begin{figure*}[t]
    \centering
    \includegraphics[width=\textwidth]{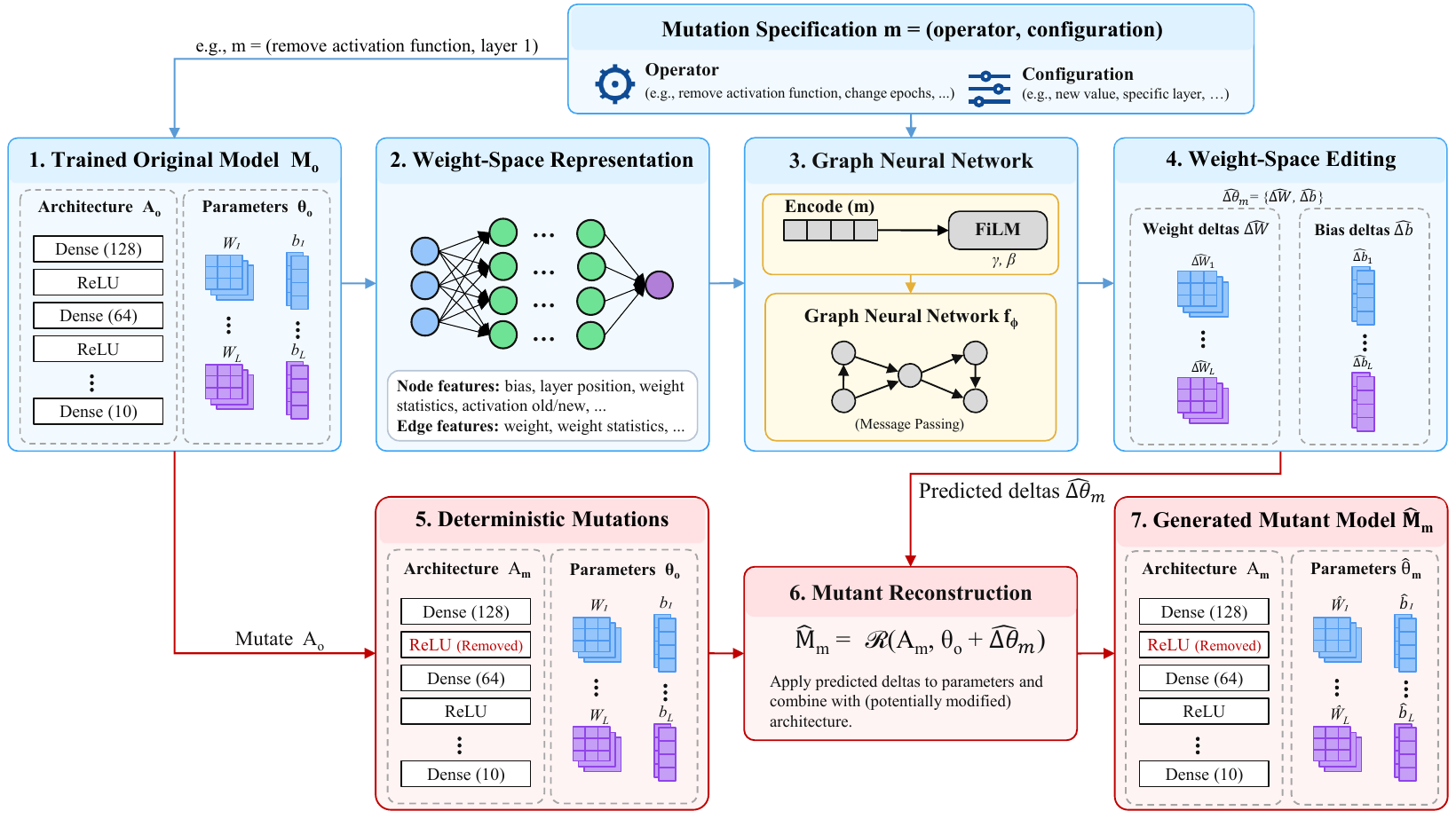}
    \caption{Mutant-generation workflow of TensorMutation.
    Given a trained original model $M_o$ and a mutation specification $m$, TensorMutation constructs a mutation-aware neural graph, uses the trained GNN $f_\phi$ to predict parameter changes, applies deterministic architectural changes when required, and reconstructs the executable mutant $\widehat{M}_m$ without retraining.}
    \label{fig:tensormutation_overview}
\end{figure*}

Fig.~\ref{fig:tensormutation_overview} summarizes the complete mutant-generation process of TensorMutation.
TensorMutation takes the trained original model $M_o$ and the mutation specification $m$ as input (Stage~1).
It next transforms $M_o$ into the mutation-aware neural graph described in Section~\ref{sec:neural-graph} (Stage~2).
The graph carries local mutation information where applicable, while the mutation specification $m$ is supplied separately as the graph-level condition.
The neural graph is then processed by the GNN $f_\phi$, which is conditioned on $m$ as described in Section~\ref{sec:mutation-conditioning} (Stage~3).
This conditioning allows the same GNN to handle different mutation operators and configurations.
Instead of predicting the complete set of mutant parameters, the GNN predicts the mutation-induced parameter change $\widehat{\Delta\theta}_m$ defined in~\eqref{eq:predicted_delta} (Stage~4).
For operators that require an explicit architectural change, TensorMutation applies the deterministic mutation mechanics described in Section~\ref{sec:deterministic-mutation-mechanics} (Stage~5).
These mechanics modify the architecture as required by $m$ while retaining the original parameters as the base for the predicted edits.
If no structural change is required, the original architecture is retained.
Stage~6 combines the predicted parameter changes with the architecture produced by the deterministic branch.
TensorMutation then reconstructs the executable mutant.
The resulting mutant $\widehat{M}_m$ is shown in Stage~7.

\subsection{Training Corpus Construction}
\label{sec:training-corpus-construction}

The GNN of TensorMutation is trained on pairs of original models and their pre-training mutants.
For each original model, a mutation operator and configuration are applied to its training program, which is then trained to obtain the mutant.
The resulting pair, together with its specification $m$, forms one training tuple $(M_o, m, M_m)$ and captures the effect of a known pre-training mutation.
A key requirement is neuron-to-neuron correspondence between the two models. 
Hidden neurons can be permuted without necessarily changing the function represented by a network.
Direct comparison of independently ordered models could therefore confuse the mutation effect with arbitrary neuron permutations. 
We preserve positional correspondence between the original and mutant models and verify it before the pair is used for learning (Section~\ref{sec:corpus-preparation}).
Thus, a weight $w_{ij}$ or bias $b_j$ at a given position refers to the same connection or neuron in both models,
\begin{equation}
    w_{ij}^{o} \leftrightarrow w_{ij}^{m},
    \qquad
    b_j^{o} \leftrightarrow b_j^{m},
    \label{eq:alignment}
\end{equation}
where superscripts $o$ and $m$ denote the original and mutated models, respectively. 
This correspondence provides a valid basis for constructing the supervision targets for weight and bias deltas.

\subsection{Neural-Graph Representation}
\label{sec:neural-graph}

Each aligned original--mutant pair is converted into one supervised neural-graph sample. Following the neural-graph formulation of Kofinas et al.~\cite{kofinas_graph_2024}, the original model provides the graph structure and input features. 
Neurons are represented as nodes and trainable connections as directed edges. 
The corresponding mutant is used to construct the supervision targets.
For a node $v_j$ and an edge $e_{ij}$ from node $v_i$ to node $v_j$, these targets are the mutation-induced parameter differences
\begin{equation}
    \Delta b_j = b_j^{m} - b_j^{o},
    \qquad
    \Delta w_{ij} = w_{ij}^{m} - w_{ij}^{o}.
\end{equation}
For bias removal, the absent mutant bias is taken as zero, so $\Delta b_j = -b_j^o$.
Each node is therefore associated with an input feature vector $\mathbf{x}_j$ and a target $\Delta b_j$, while each edge has an input feature vector $\mathbf{x}_{ij}$ and a target $\Delta w_{ij}$.
Fig.~\ref{fig:neural_graph} shows one such sample together with its mutation specification $m$, whose conditioning of the GNN through FiLM is described in Section~\ref{sec:mutation-conditioning}.

The node representation has 21 dimensions and combines original-model information with localized mutation information.
It includes the bias value, layer position and width, incoming and outgoing weight statistics, activation identities before and after the requested mutation, bias and dropout information, and an indicator identifying the targeted layer. 
The 10-dimensional edge representation includes the original weight, normalized weight statistics, its position within the layer weight matrix, layer position, and adjacent-layer width information. 
The feature dimensions remain fixed while the numbers of nodes and edges vary with the network architecture, allowing the same GNN to process models of different depths and widths. 

\subsection{Mutation Conditioning}
\label{sec:mutation-conditioning}

Each neural-graph sample is associated with the mutation specification \(m\) defined in Section~\ref{sec:problem-formulation-and-overview}.
The graph representation captures the local information needed to identify where the requested mutation acts.
In parallel, the mutation specification provides a graph-level condition that tells the GNN which transformation to predict.
TensorMutation encodes the mutation specification into a conditioning vector $z_m$.
The encoding contains a learned operator representation together with operator-specific configuration information.
For numeric mutations, the configuration captures the change in the requested value.
Categorical mutations encode the requested target value, while activation mutations encode the activation transition.

\begin{figure}[t]
    \centering
    \includegraphics[width=\columnwidth]{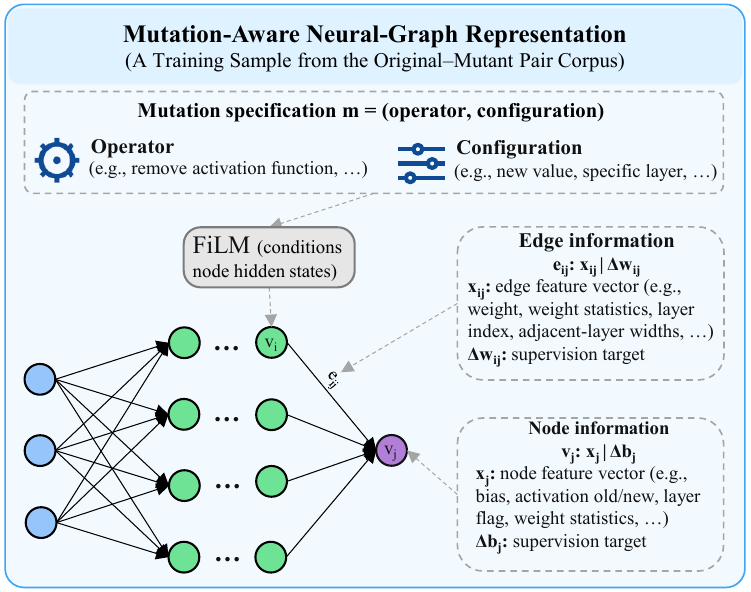}
    \caption{Mutation-aware neural-graph representation used by TensorMutation.
    Each original--mutant pair yields one supervised graph sample.
    The original model provides the node and edge input features, while the
    corresponding parameter differences provide the bias- and weight-delta
    supervision targets. The mutation specification is associated with the
    graph and conditions GNN processing through FiLM.}
    \label{fig:neural_graph}
\end{figure}

The conditioning vector is applied through FiLM~\cite{perez_film_2018}.
It maps $z_m$ to affine parameters $\gamma(z_m)$ and $\beta(z_m)$ and uses them to modulate the hidden state of each node,
\begin{equation}
    \mathbf{h}'_j =
    \gamma(z_m)\odot\mathbf{h}_j + \beta(z_m).
\end{equation}
The same graph-level condition is therefore applied to all nodes belonging to the sample, while their individual graph features preserve the local structure and state of the original model.
The learned operator representation is also provided directly to the node, message-passing, update, and edge-prediction computations, providing operator identity independently of the FiLM modulation.
For readability, Fig.~\ref{fig:neural_graph} draws the FiLM block for only one node.
The same modulation is applied to every node of the sample.

\subsection{Delta-Prediction Model}
\label{sec:delta-prediction-model}

TensorMutation uses a message-passing GNN $f_\phi$~\cite{gilmer_neural_2017} to predict the parameter changes associated with the requested mutation. 
Node features are first mapped to hidden representations. 
The GNN then performs two message-passing steps with a hidden dimension of 128. 
For each edge, separate forward and reverse messages are computed and aggregated at the corresponding nodes. 
This bidirectional propagation allows each node representation to incorporate information from both its incoming and outgoing neighborhoods. 
The mutation condition modulates the node hidden states through FiLM as described above.

After message passing, separate node and edge heads produce the parameter deltas. 
The node head predicts a bias delta $\widehat{\Delta b}_j$ for each node $v_j$. 
The edge head combines the conditioned hidden representations of the source and destination nodes with the edge representation to predict a weight delta $\widehat{\Delta w}_{ij}$ for each edge $e_{ij}$. 
FiLM conditioning therefore also influences the weight predictions through the conditioned endpoint representations. 
Together, the node- and edge-level predictions form the parameter transformation
\begin{equation}
    \widehat{\Delta\theta}_m =
    \left\{
        \widehat{\Delta W}_m,
        \widehat{\Delta b}_m
    \right\}.
    \label{eq:predicted_delta}
\end{equation}

The predicted edge residuals are rescaled using the root-mean-square magnitude of the corresponding source-layer weights, while bias deltas are predicted directly without layer-wise scaling.

\subsection{Learning Objective}
\label{sec:learning-objective}

TensorMutation is trained to reproduce the behavioral effect of the corresponding pre-training mutant on a fixed set of inputs, which we call probes.
For each training subject, the probes are a fixed sample of up to 512 inputs drawn once from the subject's test split.
The same probes are used for the original model and all of its mutants to provide behavioral supervision during GNN training.
For each training sample, the predicted deltas are applied to the original model and the resulting mutant is evaluated differentiably on the probes.
The behavioral objective combines two signals.
First, the output distribution of the predicted mutant is matched to that of the reference mutant using the Kullback--Leibler (KL) divergence~\cite{kullback_information_1951} over all probes.
Second, on the probes that the original model classifies correctly, the predicted mutant is trained to reproduce the reference mutant's kill behavior.
A small regularization term discourages unnecessarily large parameter changes.
The behavioral objective is
\begin{equation}
    \mathcal{L}_{\mathrm{beh}}
    =
    \mathcal{L}_{\mathrm{KL}}
    +
    \mathcal{L}_{\mathrm{kill}}
    +
    \lambda_r \mathcal{L}_{\mathrm{reg}},
    \label{eq:behavioral-loss}
\end{equation}
where each $\mathcal{L}$ denotes a loss term minimized during training, $\mathcal{L}_{\mathrm{reg}}$ is an $L_2$ penalty on the predicted weight and bias deltas, and $\lambda_r$ controls its contribution.

Behavioral agreement alone does not require the predicted and reference mutants to change their parameters in the same way.
The objective therefore also includes direct weight-space supervision.
It aligns the directions of the predicted and reference parameter deltas,
\begin{align}
    \mathcal{L}_{W} &= 1-\cos\!\left(\widehat{\Delta W}_m,\,\Delta W_m\right), \\
    \mathcal{L}_{b} &= 1-\cos\!\left(\widehat{\Delta b}_m,\,\Delta b_m\right).
\end{align}
The full objective is
\begin{equation}
\label{eq:full-loss}
    \mathcal{L}
    =
    \mathcal{L}_{\mathrm{beh}}
    +
    \lambda_W \mathcal{L}_{W}
    +
    \lambda_b \mathcal{L}_{b},
\end{equation}
where $\lambda_W$ and $\lambda_b$ control the contributions of the weight- and bias-delta directional terms, respectively.

When the reference delta is non-zero, the weight-space terms constrain the direction of the predicted delta, but not its magnitude.
The scale of a predicted weight delta is adjusted by the layer-wise rescaling described above, while its magnitude is otherwise shaped by the behavioral objective and regularization.
When a reference delta is zero, its direction is undefined.
The corresponding weight-space term therefore penalizes a non-zero prediction instead of applying cosine alignment.
Separately, we retain no-op training pairs whose target parameter transformation is zero.
These pairs serve as identity anchors, allowing the model to learn the conditional zero transformation.

\subsection{Training the GNN}
\label{sec:training-gnn}

\begin{figure}[!t]
    \centering
    \includegraphics[width=\columnwidth]{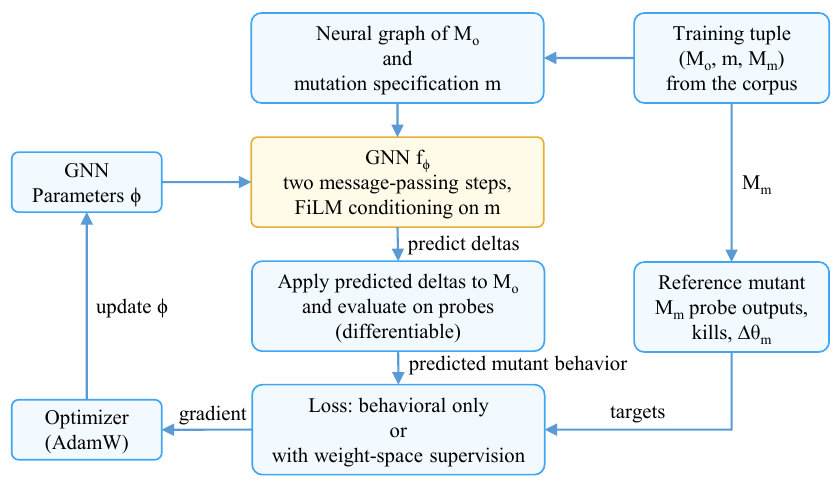}
    \caption{Offline training of the TensorMutation GNN.
    The trained GNN is subsequently reused unchanged in the mutant-generation workflow of Fig.~\ref{fig:tensormutation_overview}.}
    \label{fig:tm-training}
\end{figure}

The components described above form the GNN training loop shown in Fig.~\ref{fig:tm-training}.
Each GNN variant is trained once on the training partition of the corpus (Section~\ref{sec:corpus-preparation}).
For each training tuple $(M_o,m,M_m)$, the GNN receives the neural graph of the original model $M_o$ together with the mutation specification $m$.
The reference mutant $M_m$ is used only to provide the training targets.
Its probe outputs and kill behavior provide the behavioral targets, while its parameter differences from $M_o$ provide the weight-space targets when these are used.
The GNN predicts the parameter deltas $\widehat{\Delta\theta}_m$.
These deltas are added to the original parameters, and the resulting model is evaluated differentiably on the probes.
The loss is then backpropagated to update the GNN parameters $\phi$.

The GNN can be trained with the behavioral objective in~\eqref{eq:behavioral-loss} alone or with the full objective in~\eqref{eq:full-loss}.
The variants evaluated in this study, together with the optimization and checkpoint-selection settings, are described in Section~\ref{sec:model-configuration}.
After training, the selected GNN is fixed and used for mutant generation as shown in Fig.~\ref{fig:tensormutation_overview}.

\subsection{Deterministic Mutation Mechanics}
\label{sec:deterministic-mutation-mechanics}

Some mutation requests contain an inference-visible change that is known directly from the mutation specification.
TensorMutation applies such changes deterministically rather than requiring the GNN to infer them.
This is required for the activation operators.
Changing an activation function replaces the activation of the targeted layer with the requested one.
Removing an activation function replaces it with the identity function, implemented as a linear activation.
When the requested activation already matches the effective activation of the targeted layer, the request is a no-op.
In this case, the architecture is left unchanged and the predicted parameter deltas are set to zero, which guarantees an exact identity transformation rather than relying on the learned approximation.
For the remaining operators, reconstruction retains the original architecture.
Training-configuration mutations and bias removal are represented entirely through the predicted parameter deltas.

\subsection{Mutant Reconstruction}
\label{sec:mutant-reconstruction}

After the predicted parameter changes and any required architectural modifications are available, TensorMutation combines them to construct the executable mutant.
The reconstruction corresponds to Stages~6 and~7 of Fig.~\ref{fig:tensormutation_overview}.
Formally,
\begin{equation}
    \widehat{M}_m =
    \mathcal{R}\!\left(
        A_m,\,
        \theta_o + \widehat{\Delta\theta}_m
    \right),
    \label{eq:reconstruction}
\end{equation}
where $\mathcal{R}(\cdot)$ denotes model reconstruction.
The resulting mutant parameters are $\widehat{\theta}_m=\theta_o+\widehat{\Delta\theta}_m$, giving the executable mutant $\widehat{M}_m=(A_m,\widehat{\theta}_m)$.

The same reconstruction procedure is also used to generate two auxiliary variants for controlled evaluation.
The \emph{oracle} replaces the predicted deltas with the observed parameter deltas $\Delta\theta_m$ of the corresponding pre-training mutant.
The \emph{norm-matched random} variant replaces the learned deltas with random directions while matching the node- and edge-delta norms separately.
Their roles in the evaluation are described in Section~\ref{sec:comparison-methods}.

\section{Study Design}
\label{sec:study-design}

\begin{figure*}[!t]
    \centering
    \includegraphics[width=\textwidth]{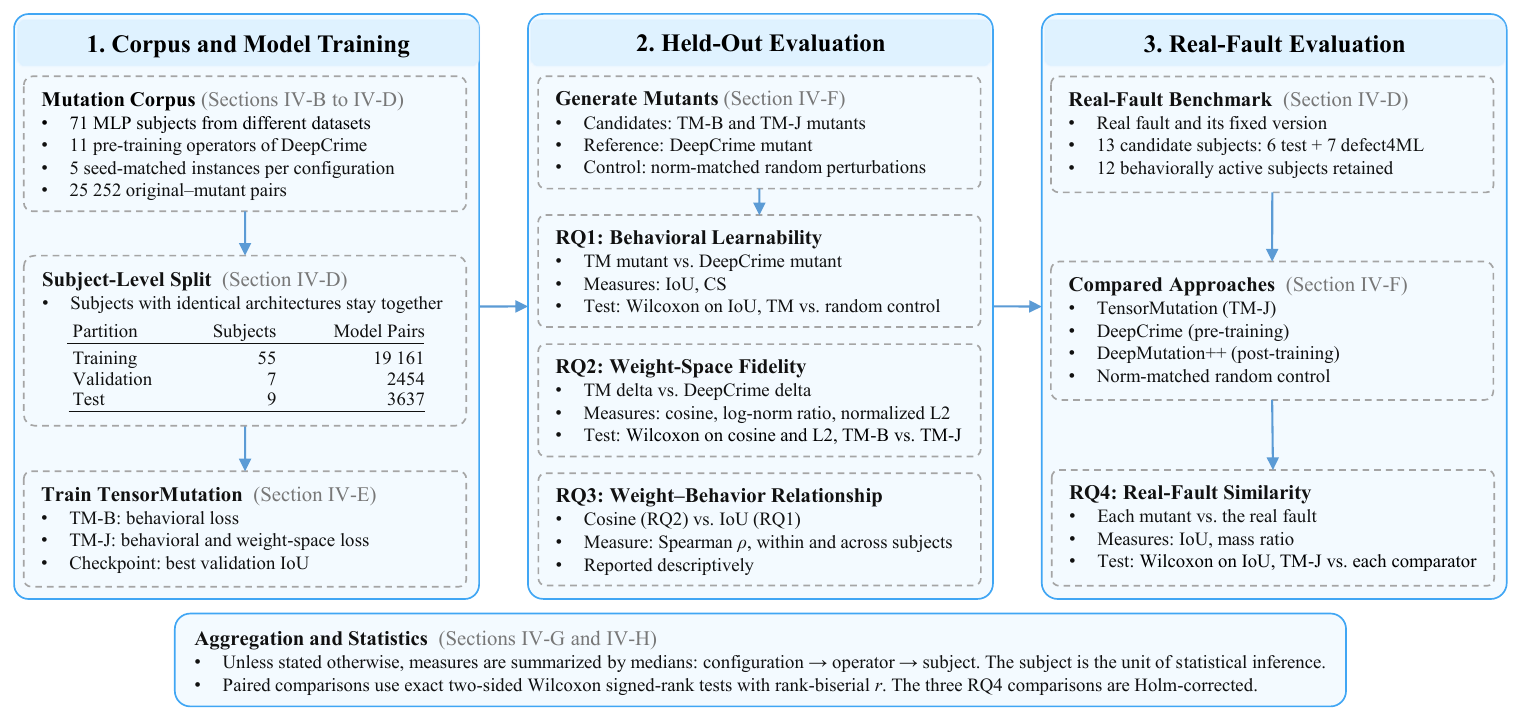}
    \caption{Overview of the TensorMutation study design.
    The study constructs and splits the mutation corpus, trains TensorMutation, evaluates behavioral and weight-space fidelity on held-out subjects, and compares the generated mutants with real faults.
    TM-B and TM-J denote TensorMutation trained with the behavioral objective only and with the full objective, respectively (Section~\ref{sec:model-configuration}).}
    \label{fig:study-design-overview}
\end{figure*}

\noindent In this section, we describe how we set up our experiments to evaluate TensorMutation.
We present the RQs, subject collection, mutation operators, corpus preparation, model configuration, comparison methods, evaluation measures, statistical analysis, and design ablations.
Fig.~\ref{fig:study-design-overview} summarizes the experimental design used to evaluate TensorMutation.

\subsection{Research Questions}
\label{sec:research-questions}

To evaluate the learnability and effectiveness of TensorMutation, we organize the study around four RQs.

\noindent\textbf{RQ1 (Behavioral Learnability):}
\emph{Can the behavioral effects of pre-training mutations be learned as post-training weight-space transformations that generalize to unseen subjects?}

RQ1 tests the central idea behind TensorMutation: whether a learned parameter transformation can reproduce the behavior of the corresponding pre-training mutant on subjects not seen during training.
On held-out subjects, we measure the kill-profile similarity of each TensorMutation mutant to its corresponding DeepCrime mutant.
We compare these scores with those of norm-matched random counterparts, which retain the magnitude of the TensorMutation prediction but randomize its direction.
This comparison tests whether the learned direction contributes to behavioral fidelity beyond perturbation magnitude alone.

\noindent\textbf{RQ2 (Weight-Space Fidelity):}
\emph{How faithfully does TensorMutation reproduce mutation-induced parameter changes, and does explicit weight-space supervision improve this fidelity?}

RQ2 examines how closely the predicted parameter changes match those produced by the corresponding DeepCrime mutant on held-out subjects.
We measure their agreement in direction, magnitude, and normalized error using cosine similarity, log-norm ratio, and normalized $L_2$ distance.
We compare training with the behavioral objective in~\eqref{eq:behavioral-loss}
against the full objective in~\eqref{eq:full-loss}, which adds direct weight-space supervision.
This shows whether explicit weight-space supervision improves parameter fidelity and whether it affects behavioral fidelity.

\noindent\textbf{RQ3 (Weight--Behavior Relationship):}
\emph{How does weight-space fidelity relate to behavioral fidelity, and does this relationship differ within and across subjects?}

This question asks whether more accurate parameter transformations also lead to more similar mutant behavior.
We therefore relate the weight-space fidelity measures from RQ2 to the behavioral fidelity measured in RQ1.
The relationship is examined using Spearman correlation both within subjects and across subjects.
This distinction shows whether the association holds only among mutations of the same model or also across different models.

\noindent\textbf{RQ4 (Real-Fault Similarity):}
\emph{To what extent does the behavior of TensorMutation mutants resemble that of real faults, and how does this resemblance compare with existing mutation approaches and controls?}

This question moves beyond reproducing pre-training mutants and asks whether TensorMutation mutants resemble the behavior of real faults.
We evaluate TensorMutation on the 12 behaviorally active real-fault subjects defined in Section~\ref{sec:corpus-preparation} and measure the kill-profile similarity of each mutant to the corresponding real fault.
TensorMutation is compared with DeepCrime, DeepMutation++, and the norm-matched random control, whose roles are described in Section~\ref{sec:comparison-methods}.
We further analyze behavioral severity and direction to explain the observed differences.

\subsection{Subject Collection}
\label{sec:subject-collection}

Our study focuses on MLP classifiers implemented in TensorFlow/Keras.
We build the development corpus from subjects previously reproduced in our empirical study of mutant realism~\cite{ahmed_empirical_2025} and supplement them with a new collection of MLP subjects.
The previous study reproduced real faults from CleanML~\cite{li_cleanml_2021}, DeepFD~\cite{cao_deepfd_2022}, DeepLocalize~\cite{wardat_deeplocalize_2021}, and defect4ML~\cite{morovati_bugs_2023}.
For TensorMutation development, we consider the reproduced subjects from CleanML, DeepFD, and DeepLocalize whose models are MLP classifiers.
We further exclude three subjects whose neural graphs exceed one million edges.
Their graphs contain between 6.9 and 12.7 million edges, more than ten times the largest retained subject (0.67 million), and exceeded the memory and time budget of GNN training.
This results in 19 CleanML, 17 DeepFD, and 10 DeepLocalize subjects.
The fixed version of each retained subject provides the original model from which pre-training mutants are generated.

To increase the diversity available for cross-subject learning, we additionally curated TFEX-MLP (TensorFlow Examples of MLPs), a set of 25 MLP classifiers.
Its subjects are source-traceable TensorFlow/Keras programs collected from independently authored tutorials, books, repositories, and other public resources.
The original architecture and training semantics are retained as closely as possible, while fixed data partitions and compatibility adaptations support reproducible execution and mutation generation.
The resulting development corpus contains 71 subjects.
The eligible defect4ML subjects serve only as an additional real-fault benchmark for RQ4.

\subsection{Mutation Operators}
\label{sec:mutation-operators}

We generate the mutation corpus with the pre-training mutation tool DeepCrime~\cite{humbatova_deepcrime_2021}.
It provides 24 mutation operators for both the training program and the training data.
These operators were derived from three studies of real faults~\cite{humbatova_taxonomy_2020, zhang_empirical_2018, islam_comprehensive_2019}.
This study focuses on the training program because nearly all published datasets of real faults concern faults in the DL program.
A systematic survey~\cite{jahangirova_real_2026} identified 16 studies with such datasets, and all except CleanML concern program faults.
We selected the 11 DeepCrime operators that mutate the training program, preserve the neuron correspondence of~\eqref{eq:alignment}, and apply to our subjects.
We excluded eight of DeepCrime's 19 training-program operators for three reasons.
Changing the weight initialization was excluded because training then starts from an independent draw of weights, so the correspondence in~\eqref{eq:alignment} no longer holds.
Removing the validation set was excluded because it left the trained parameters unchanged on 13 of the 17 subjects where it applied, so it offers few nontrivial training pairs.
The other six were excluded because they applied to fewer than ten or to none of our subjects: adding an activation function, changing or removing weight regularization, changing the early-stopping patience, adding a bias, and changing gradient clipping.

Table~\ref{tab:mutation-operators} lists the selected operators with the fault type each one emulates and the configuration space used to generate mutants.
The configuration spaces follow DeepCrime's defaults.
\begin{table*}[t]
\centering
\caption{Selected DeepCrime Mutation Operators with Fault Types and Configuration Spaces}
\label{tab:mutation-operators}
\footnotesize
\begin{tabularx}{\textwidth}{@{}l l l X@{}}
\toprule
ID & Operator & Fault type & Configuration space \\
\midrule
HBS & Change batch size & Suboptimal batch size & $\{b/4,\ b/2,\ 2b,\ 4b\}$ \\
HLR & Decrease learning rate & Suboptimal learning rate & $(\sim\!0,\ lr]$, binary search with precision $lr/10$ \\
HNE & Change number of epochs & Suboptimal number of epochs & $[1,\ e]$, binary search with precision $e/10$ \\
HDB & Disable data batching & Data batching required & Batch size set to the training-set size; only if batching is used \\
ACH & Change activation function & Wrong type of activation function & Each layer with a non-linear activation $\times$ each other non-linear activation \\
ARM & Remove activation function & Missing activation function & Each layer with a non-linear activation \\
RAW & Add weights regularization & Excess weights regularization$^\dagger$ & Each layer without regularization $\times$ \{L1, L2, L1L2\} \\
RCD & Change dropout rate & Excessively high/low dropout rate$^\dagger$ & Each dropout layer $\times$ four rates from $(0, 1]$ excluding the original value \\
WRB & Remove bias from a layer & Bias needed in a layer & Each layer with a bias \\
LCH & Change loss function & Wrong selection of loss function & Each loss other than the original that is compatible with the label encoding \\
OCH & Change optimization function & Wrong optimization function & Each optimizer other than the original \\
\bottomrule
\multicolumn{4}{@{}p{\textwidth}@{}}{\footnotesize $b$, $lr$, and $e$ denote the subject's original batch size, learning rate, and number of epochs.
Activation, loss, and optimizer candidates are taken from DeepCrime's default Keras lists~\cite{humbatova_deepcrime_2021}.
Fault types are leaves of the taxonomy of Humbatova et al.~\cite{humbatova_taxonomy_2020}.
$^\dagger$RAW and RCD were derived from Zhang et al.~\cite{zhang_empirical_2018} and Islam et al.~\cite{islam_comprehensive_2019}, respectively. As these studies define no taxonomy leaves, their fault type labels follow Xie et al.~\cite{xie_mutants_2026}.}
\end{tabularx}
\end{table*}
The selected operators target fault types frequently observed in real DL programs.
Among the 67 program faults reproduced in prior work~\cite{ahmed_empirical_2025}, 54 (80.60\%) contain at least one fault type emulated by a selected operator.
Cao et al.~\cite{cao_deepfd_2022} reported that activation function, number of epochs, learning rate, loss function, and optimizer faults collectively account for 73.55\% of the fault types recorded in their benchmark, counting each fault type once per program.

\subsection{Corpus Preparation}
\label{sec:corpus-preparation}

\paragraph{Mutation corpus}
For each subject, we train five instances of the original model and generate the applicable DeepCrime mutants from the configurations in Table~\ref{tab:mutation-operators}.
Original and mutant programs use matched per-instance seeds and deterministic GPU operations.
Each mutant is paired with its seed-matched original, and pairs with non-finite parameters are discarded.
Before constructing the parameter-delta targets, we check the saved models for pair completeness and positional tensor-shape compatibility, accounting for expected structural changes such as bias removal.
Identity mutants, which request the model's existing configuration, produce exactly zero deltas under this setup.
To prevent architectural overlap across partitions, subjects with identical architectures are grouped before splitting.
The split was fixed before TensorMutation training and yields 55 training, seven validation, and nine held-out test subjects, with 19\,161, 2454, and 3637 retained original--mutant pairs, respectively.
Test subjects, including their test inputs, are excluded from TensorMutation training and model selection, and validation subjects are used only for checkpoint selection.
Only the training partition is used to construct the vocabularies required by TensorMutation.

\paragraph{Real-fault benchmark}
Real faults are used only for evaluation.
The benchmark contains 13 subjects with a real fault and its fixed version: the six test subjects from CleanML, DeepFD, and DeepLocalize, and seven MLP classifiers from defect4ML.
A subject is included only if at least one input kills the real fault in three or more of its five seed-matched instances (Section~\ref{sec:evaluation-measures}).
This yields 12 subjects, which we call behaviorally active.

\subsection{Model Configuration}
\label{sec:model-configuration}

The GNN model uses a hidden dimension of 128 and two message-passing steps.
Training uses AdamW with a learning rate of $10^{-3}$ and weight decay of $10^{-4}$ for up to 200 epochs.
The behavioral loss uses up to 512 probe inputs per subject, with $\lambda_r=10^{-3}$.
We evaluate the validation set every five epochs and retain the checkpoint with the highest validation IoU on the probes (Section~\ref{sec:evaluation-measures}).
We train two otherwise identical variants: TensorMutation-behavioral (TM-B) uses only the behavioral loss~\eqref{eq:behavioral-loss}, whereas TensorMutation-joint (TM-J) adds direct weight-space supervision~\eqref{eq:full-loss} with $\lambda_W=\lambda_b=0.1$.
TM-J is the primary model and the one evaluated against real faults in RQ4.

\subsection{Comparison Methods}
\label{sec:comparison-methods}

TensorMutation is compared with three references, each serving a distinct role.
DeepCrime~\cite{humbatova_deepcrime_2021} is the pre-training reference: its mutants provide the target transformations for RQ1 to RQ3 and a comparator against real faults in RQ4.
We use DeepMutation++~\cite{hu_deepmutation_2019} as the manually defined post-training comparator because it has a public implementation and is peer reviewed, whereas MuNN~\cite{shen_munn_2018} has no public implementation and MuFF~\cite{kim_muff_2025} was not peer reviewed at the time of this study.
For each subject, we apply the applicable DeepMutation++ feedforward operators using the default configuration and a relative accuracy threshold of 0.9.
We generate five mutant variants per operator and original instance and keep only complete groups, so some DeepMutation++ comparisons use fewer than five instances.
DeepMutation++ therefore differs from TensorMutation in operator coverage and configuration granularity, so RQ4 compares real-fault similarity across approaches.
Norm-matched random perturbations serve as the direction control.
For each TensorMutation prediction, we generate three perturbations that separately match the predicted node- and edge-delta norms while randomizing their directions.
All comparisons use the same seed-matched original model instances.
As an integrity check, we reconstruct oracle mutants by applying the true DeepCrime deltas through the TensorMutation reconstruction pipeline.
The pre-specified integrity criterion requires their kill indicators to match those of the corresponding DeepCrime mutants exactly.
The criterion held for all 3637 test pairs and for 4449 of the 4450 real-fault pairs.
The single exception arises from a numerical near-tie between the two highest class scores on one input.

\subsection{Evaluation Measures}
\label{sec:evaluation-measures}

\paragraph{Behavioral measures}
All behavioral measures use the full test split $T$ of each subject and follow the definitions of~\cite{ahmed_empirical_2025}.
An input is a killing input (KI) for a model instance if its seed-matched original classifies it correctly and the instance misclassifies it~\cite{humbatova_deepcrime_2021}.
The killing probability $KP(t)$ of input $t$ is the fraction of the available seed-matched instances for which $t$ is a killing input.
The behavioral mass of a candidate is $\sum_{t \in T} KP(t)$, and its mass ratio is this mass divided by that of the reference, computed per configuration and summarized hierarchically like all other measures.
For DeepMutation++, which yields five variants per original instance, we form five evaluation sets by grouping the $j$-th variant from each seed-matched original.
We compute the measures for each set and average them.
The three norm-matched random perturbations are handled analogously.

We measure the behavioral similarity between a candidate $C$ and a reference $R$ as
\begin{equation}
\mathrm{IoU} =
\frac{\sum_{t \in T} \min(KP_C(t), KP_R(t))}
     {\sum_{t \in T} \max(KP_C(t), KP_R(t))},
\label{eq:iou}
\end{equation}
where the reference is the corresponding DeepCrime configuration in RQ1 and the real fault in RQ4.
IoU is bounded by the ratio of the smaller to the larger behavioral mass, so it penalizes both missed and excess kills.
We report coupling strength (CS), $\sum_{t \in T}\min(KP_C(t),KP_R(t))/\sum_{t \in T} KP_C(t)$, with CS defined as zero when the candidate mass is zero.
We use CS descriptively but not as the primary cross-method measure because its interpretation depends on the candidate's behavioral mass.
We therefore use \emph{coupling} only for this directional measure and \emph{similarity} for IoU.

\paragraph{Weight-space measures}
For RQ2, we compare the predicted delta $\widehat{\Delta\theta}$ with the DeepCrime delta $\Delta\theta$ over all aligned weights and biases, using cosine similarity for direction, the log-norm ratio $\log_{10}(\|\widehat{\Delta\theta}\|/\|\Delta\theta\|)$ for magnitude, and the normalized $L_2$ distance $\|\widehat{\Delta\theta}-\Delta\theta\|/\|\Delta\theta\|$ for overall error.

\paragraph{Weight--behavior relationship}
For RQ3, we use Spearman rank correlation to relate weight-space cosine similarity (RQ2) to behavioral IoU (RQ1).
We examine this relationship within subjects at the configuration level and across subjects using subject-level summaries.

Unless stated otherwise, measures are summarized hierarchically: configuration values are aggregated by the median within each operator, operator medians by the median within each subject, and subject medians by the median across subjects.
Configurations whose reference has zero behavioral mass are excluded from the reported behavioral summaries, and configurations with a zero reference delta are excluded from weight-space summaries.

\subsection{Statistical Analysis}
\label{sec:statistical-analysis}

The subject is the unit of statistical inference because configurations and operators are repeated observations nested within a subject rather than independent samples.
For paired subject-level comparisons, we use two-sided exact Wilcoxon signed-rank tests on the hierarchical subject summaries, omitting zero differences, and report the rank-biserial correlation~\cite{kerby_simple_2014} as the effect size.

For RQ1, we compare TM-B and TM-J with their corresponding norm-matched random controls using subject-level IoU summaries across the nine test subjects.
For RQ2, TM-B and TM-J are paired across the same nine subjects using subject-level cosine similarity and normalized $L_2$ error.
For RQ3, we report the relationship descriptively: the median of the per-subject Spearman correlations at the configuration level for the within-subject analysis, and the Spearman correlation over the nine subject summaries for the across-subject analysis.
For RQ4, TM-J is paired separately with DeepCrime, DeepMutation++, and the norm-matched random control using subject-level IoU over the 12 real-fault subjects.
These three comparisons form the family of primary RQ4 tests, and we control its family-wise error rate with the Holm correction~\cite{holm_simple_1979} at $\alpha=0.05$.

\subsection{Design Ablations}
\label{sec:design-ablations}

We examine whether source-training context improves cross-subject transfer.
The motivation is that the effect of a mutation may depend on the source setting; for example, halving a learning rate of 0.1 differs from halving one of 0.001.
TM-J conditions on the trained neural graph and requested mutation only, and we construct two context-augmented variants.
\emph{TM-J-Narrow} additionally receives only the source value associated with the requested training-setting mutation.
\emph{TM-J-Full} receives the available source learning rate, number of epochs, batch size, optimizer, and loss function for every mutation.
Both variants inject this context through FiLM using the same context width.
All other model, training, split, and evaluation settings are kept consistent with TM-J.
We treat this comparison as exploratory rather than an additional confirmatory RQ.

\section{Results}
\label{sec:results}

\noindent This section reports the results for the four RQs and the source-context design ablation.
All values are hierarchical summaries as defined in Section~\ref{sec:evaluation-measures}, and reported values are medians unless stated otherwise.

\subsection{RQ1: Behavioral Learnability}
\label{sec:results-rq1}

Table~\ref{tab:rq1} reports the behavioral similarity between TensorMutation and the corresponding DeepCrime mutants on the nine test subjects.
TM-B achieves a median IoU of 0.216 and TM-J a median of 0.232.
Subject-level IoU ranges from 0.081 to 0.438 for TM-B and from 0.060 to 0.380 for TM-J.
Both variants therefore reproduce part of the behavioral effect of pre-training mutations on unseen subjects, although fidelity varies substantially across subjects.
TM-B and TM-J mutants have greater behavioral mass than the corresponding DeepCrime mutants on six and seven of the nine test subjects, respectively.
The median candidate-to-reference behavioral-mass ratio is 1.850 for TM-B and 1.366 for TM-J.
\begin{table}
\centering
\caption{Behavioral similarity to the corresponding DeepCrime mutants on the test subjects}
\label{tab:rq1}
\footnotesize
\setlength{\tabcolsep}{3pt}
\begin{tabular}{@{}lcccccc@{}}
\toprule
 & \multicolumn{4}{c}{IoU} & \multicolumn{2}{c}{CS} \\
\cmidrule(lr){2-5}\cmidrule(lr){6-7}
Subject & TM-B & Rand.\ B & TM-J & Rand.\ J & TM-B & TM-J \\
\midrule
cleanml018      & 0.246 & 0.153 & 0.232 & 0.118 & 0.483 & 0.483 \\
cleanml019      & 0.297 & 0.243 & 0.299 & 0.214 & 0.482 & 0.441 \\
deepfd030       & 0.162 & 0.190 & 0.179 & 0.202 & 0.163 & 0.229 \\
deepfd050       & 0.178 & 0.164 & 0.248 & 0.087 & 0.228 & 0.316 \\
deeplocalize030 & 0.091 & 0.109 & 0.099 & 0.085 & 0.355 & 0.411 \\
deeplocalize039 & 0.216 & 0.186 & 0.235 & 0.094 & 0.284 & 0.340 \\
tfex027         & 0.247 & 0.258 & 0.124 & 0.295 & 0.274 & 0.181 \\
tfex028         & 0.438 & 0.406 & 0.380 & 0.411 & 0.495 & 0.494 \\
tfex029         & 0.081 & 0.192 & 0.060 & 0.126 & 0.084 & 0.061 \\
\midrule
Median          & 0.216 & 0.190 & 0.232 & 0.126 & 0.284 & 0.340 \\
\midrule
Wins/losses vs.\ control
                & \multicolumn{2}{c}{5/4}
                & \multicolumn{2}{c}{5/4} & & \\
Wilcoxon $p$
                & \multicolumn{2}{c}{0.570}
                & \multicolumn{2}{c}{0.652} & & \\
Rank-biserial $r$
                & \multicolumn{2}{c}{0.244}
                & \multicolumn{2}{c}{0.200} & & \\
\bottomrule
\end{tabular}
\vspace{1mm}

\parbox{\columnwidth}{\footnotesize
Rand.: norm-matched random control. CS: coupling strength (Section~\ref{sec:evaluation-measures}), reported descriptively; paired tests compare IoU only.}

\end{table}

\begin{figure}[!t]
\centering
\includegraphics[width=1\linewidth]{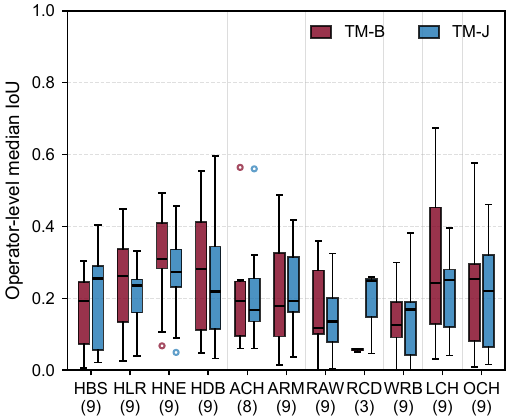}
\caption{Operator-level median IoU of TM-B and TM-J with the corresponding DeepCrime mutants across the test subjects. Numbers in parentheses give the number of test subjects to which each operator applies.}
\label{fig:rq1-operators}
\end{figure}

Fig.~\ref{fig:rq1-operators} shows the results per mutation operator.
Operator-level median IoU ranges from 0.058 (RCD) to 0.309 (HNE) for TM-B and from 0.136 (RAW) to 0.274 (HNE) for TM-J.
Changing the number of epochs achieves the highest median IoU for both variants.
TM-J achieves a higher median than TM-B on six operators and a lower one on five, so direct weight-space supervision does not change behavioral fidelity uniformly.
The norm-matched random controls achieve median IoUs of 0.190 for TM-B and 0.126 for TM-J.
Both medians favor TensorMutation, most clearly for TM-J.
However, each TensorMutation variant exceeds its corresponding control on only five of the nine subjects, and neither paired difference is statistically significant (TM-B: $p=0.570$; TM-J: $p=0.652$).
Behavioral similarity alone therefore does not establish that the learned perturbation direction is responsible for the observed fidelity.
RQ2 examines the direction directly in weight space, and RQ4 examines its contribution against real faults.

\begin{tcolorbox}[colback=gray!5, colframe=gray!60, boxrule=0.5pt, left=4pt, right=4pt, top=2pt, bottom=2pt]
\textbf{Answer to RQ1:}
TensorMutation reproduces part of the behavioral effect of pre-training mutations on unseen subjects, with median IoUs of 0.216 for TM-B and 0.232 for TM-J.
Fidelity varies across subjects and operators, and both variants usually produce greater behavioral mass than their DeepCrime references.
The norm-matched random comparison does not show a consistent subject-level advantage for the learned direction.
\end{tcolorbox}

\subsection{RQ2: Weight-Space Fidelity}
\label{sec:results-rq2}

\begin{figure}[!t]
\centering
\includegraphics[width=1\linewidth]{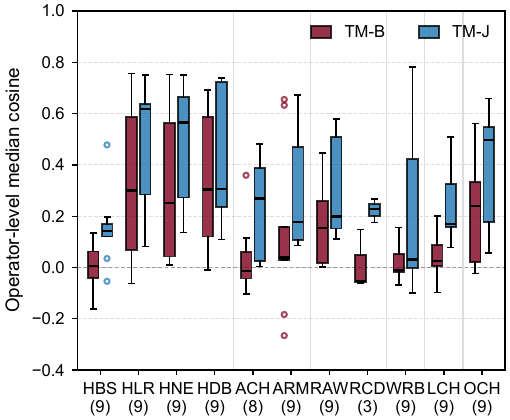}
\caption{Operator-level median cosine similarity of TM-B and TM-J with the corresponding DeepCrime deltas across the test subjects. Numbers in parentheses give the number of test subjects to which each operator applies.}
\label{fig:rq2-operators}
\end{figure}

Table~\ref{tab:rq2} compares the predicted parameter deltas with those of the corresponding DeepCrime mutants.
With the behavioral objective alone, the predicted deltas are only weakly aligned with the reference deltas, with a median cosine similarity of 0.080.
Adding direct weight-space supervision raises the median to 0.227.
TM-J improves cosine similarity on all nine subjects ($p=0.004$, rank-biserial $r=1.000$).
Fig.~\ref{fig:rq2-operators} shows that the improvement also holds across mutation operators.
TM-B's operator-level median cosine similarity is at most 0.039 for six operators and negative for three of them.
TM-J's median is at least as high as TM-B's for all 11 operators, and TM-J is higher on 84 of the 92 subject-operator combinations.

Normalized $L_2$ error also decreases on all nine subjects, from a median of 1.377 to 1.030 ($p=0.004$, rank-biserial $r=1.000$).
Since this measure divides the prediction error by the norm of the reference delta, predicting no change at all yields exactly one.
TM-B's error exceeds this level on eight of the nine subjects, whereas TM-J's median error is 1.030 and its error falls below one on four subjects.
The larger gain is therefore in direction rather than in exact reconstruction of the full delta.

Magnitude calibration shows a different pattern.
The median log-norm ratio is 0.046 for TM-B and $-0.171$ for TM-J.
TM-B's ratio is positive on five subjects and negative on four, whereas TM-J's is negative on seven.
A ratio of $-0.171$ corresponds to predicted deltas about two thirds the norm of the reference.
TM-J therefore tends to predict smaller parameter changes than DeepCrime, while TM-B shows no consistent direction of error.
Weight-space supervision does not reduce median behavioral fidelity.
TM-J has a slightly higher median IoU than TM-B (0.232 versus 0.216, Table~\ref{tab:rq1}), but it is higher on only five of the nine subjects and lower on four.
The same mixed picture holds across operators, where TM-J has the higher median on six operators and the lower one on five (Fig.~\ref{fig:rq1-operators}).
The consistent gain in direction therefore does not carry over uniformly into behavior.
RQ3 examines how directional and behavioral fidelity are related.

\begin{table}
\centering
\caption{Weight-space fidelity to the corresponding DeepCrime deltas on the test subjects}
\label{tab:rq2}
\footnotesize
\setlength{\tabcolsep}{3pt}
\begin{tabular}{@{}lcccccc@{}}
\toprule
 & \multicolumn{2}{c}{Cosine} &
   \multicolumn{2}{c}{Log-norm ratio} &
   \multicolumn{2}{c}{Norm.\ $L_2$} \\
\cmidrule(lr){2-3}\cmidrule(lr){4-5}\cmidrule(lr){6-7}
Subject & TM-B & TM-J & TM-B & TM-J & TM-B & TM-J \\
\midrule
cleanml018      & 0.146    & 0.422 &  0.046   & $-0.151$ & 1.377 & 1.030 \\
cleanml019      & 0.282    & 0.532 & $-0.114$ & $-0.226$ & 1.009 & 0.925 \\
deepfd030       & 0.020    & 0.115 &  0.130   & $-0.171$ & 1.717 & 1.125 \\
deepfd050       & 0.080    & 0.227 & $-0.175$ & $-0.479$ & 1.074 & 0.983 \\
deeplocalize030 & 0.030    & 0.156 &  0.156   & $-0.154$ & 1.777 & 1.117 \\
deeplocalize039 & 0.037    & 0.179 & $-0.111$ & $-0.371$ & 1.124 & 0.988 \\
tfex027         & 0.165    & 0.508 &  0.383   &  0.407   & 2.430 & 2.129 \\
tfex028         & $-0.004$ & 0.125 &  0.536   &  0.218   & 3.543 & 1.895 \\
tfex029         & 0.360    & 0.481 & $-0.274$ & $-0.189$ & 0.986  & 0.878 \\
\midrule
Median          & 0.080    & 0.227 &  0.046   & $-0.171$ & 1.377 & 1.030 \\
\midrule
TM-J better
                & \multicolumn{2}{c}{9/9}
                & & &
                \multicolumn{2}{c}{9/9} \\
Wilcoxon $p$
                & \multicolumn{2}{c}{0.004}
                & & &
                \multicolumn{2}{c}{0.004} \\
Rank-biserial $r$
                & \multicolumn{2}{c}{1.000}
                & & &
                \multicolumn{2}{c}{1.000} \\
\bottomrule
\end{tabular}
\vspace{1mm}

\parbox{\columnwidth}{\footnotesize
Log-norm ratio: $\log_{10}$ of the predicted-to-reference delta norm; negative values indicate smaller predicted deltas. No test is reported for it, since calibration is best at zero rather than at either extreme.}

\end{table}

\begin{tcolorbox}[colback=gray!5, colframe=gray!60, boxrule=0.5pt, left=4pt, right=4pt, top=2pt, bottom=2pt]
\textbf{Answer to RQ2:}
Direct weight-space supervision improves the direction and normalized error of the predicted deltas on every test subject, without reducing the median behavioral fidelity.
Median cosine similarity increases from 0.080 to 0.227, while normalized $L_2$ error decreases from 1.377 to 1.030, close to the error of predicting no change.
Magnitude remains imperfectly calibrated: TM-B's signed errors vary across subjects, while TM-J tends to predict smaller deltas than the reference.
\end{tcolorbox}

\subsection{RQ3: Weight--Behavior Relationship}
\label{sec:results-rq3}

We next examine whether greater directional fidelity in weight space is associated with greater behavioral fidelity.
The analysis aligns the 697 configurations listed in Table~\ref{tab:rq3} for each TensorMutation variant and relates cosine similarity to behavioral IoU using Spearman correlation.
Within subjects, the relationship is positive for most subjects.
The median of the nine subject-level correlations is $\rho=0.339$ for TM-B and $\rho=0.241$ for TM-J, and the correlation is positive on seven of the nine subjects for TM-B and on eight for TM-J.
TM-J has the lower median but the more consistent sign; its only negative correlation occurs on tfex028, which is also negative for TM-B.
Thus, within a given subject, configurations whose predicted parameter changes are more closely aligned with the corresponding DeepCrime changes tend to reproduce their behavioral effects more closely.
The associations are weak to moderate, indicating that weight-space direction is only partly associated with variation in behavioral fidelity.
The relationship does not persist across subjects.
Using the hierarchical subject summaries from RQ1 and RQ2, the correlation between subject-level cosine similarity and IoU is $\rho=-0.100$ for TM-B and $\rho=-0.083$ for TM-J.
Subjects with higher overall directional fidelity therefore do not systematically exhibit higher behavioral fidelity.

These results also clarify the effect observed in RQ2.
Although direct weight-space supervision substantially increases cosine similarity for TM-J, it does not produce a stronger configuration-level association between cosine similarity and behavioral IoU.
Improved directional reconstruction and behavioral resemblance are therefore related, but they are not interchangeable measures.

\begin{tcolorbox}[colback=gray!5, colframe=gray!60, boxrule=0.5pt, left=4pt, right=4pt, top=2pt, bottom=2pt]
\textbf{Answer to RQ3:}
Better directional fidelity in weight space tends to correspond to better behavioral fidelity within a subject, with median Spearman correlations of 0.339 for TM-B and 0.241 for TM-J.
This relationship does not persist across subjects, where the correlations are $-0.100$ and $-0.083$, respectively.
Weight-space fidelity is therefore associated with behavioral fidelity within subjects but does not explain differences in behavioral fidelity across subjects.
\end{tcolorbox}

\begin{table}[t]
\centering
\caption{Spearman correlation between configuration-level cosine similarity and IoU on the test subjects}
\label{tab:rq3}
\footnotesize
\setlength{\tabcolsep}{4pt}
\begin{tabular}{@{}lccc@{}}
\toprule
 & & \multicolumn{2}{c}{Spearman $\rho$} \\
\cmidrule(lr){3-4}
Subject & Configurations & TM-B & TM-J \\
\midrule
cleanml018      & 53  & 0.242    & 0.241 \\
cleanml019      & 54  & 0.339    & 0.202 \\
deepfd030       & 119 & 0.404    & 0.151 \\
deepfd050       & 87  & 0.360    & 0.573 \\
deeplocalize030 & 120 & 0.412    & 0.119 \\
deeplocalize039 & 55  & 0.515    & 0.552 \\
tfex027         & 67  & $-0.122$ & 0.288 \\
tfex028         & 60  & $-0.228$ & $-0.136$ \\
tfex029         & 82  & 0.038    & 0.251 \\
\midrule
Median          &     & 0.339    & 0.241 \\
Positive        &     & 7/9     & 8/9 \\
\midrule
Across subjects &     & $-0.100$ & $-0.083$ \\
\bottomrule
\end{tabular}
\vspace{1mm}

\parbox{\columnwidth}{\footnotesize
The across-subject row correlates the subject-level IoU and cosine similarity  of Tables~\ref{tab:rq1} and~\ref{tab:rq2}.}

\end{table}

\subsection{RQ4: Real-Fault Similarity}
\label{sec:results-rq4}

\begin{figure*}[t]
\centering
\includegraphics[width=\textwidth]{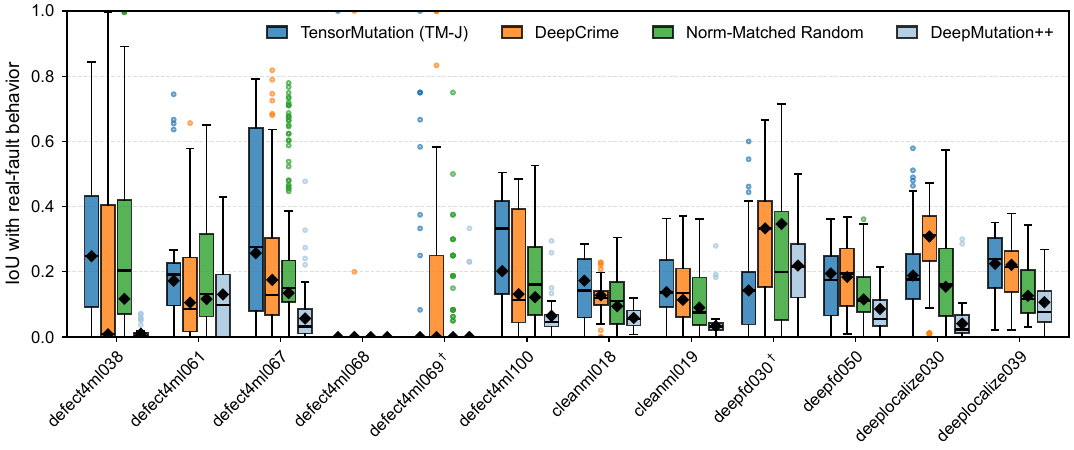}
\caption{IoU with the real fault on the 12 subjects.
Boxes show the IoU of individual mutants, and black diamonds show the subject summaries of Table~\ref{tab:rq4-subjects}, on which the tests are based.
$^\dagger$Subjects with fewer than 50 real-fault kills.}
\label{fig:rq4}
\end{figure*}

We now evaluate TM-J as the primary variant of TensorMutation, whose directional fidelity exceeds TM-B's on every test subject with no loss in median behavioral fidelity (Section~\ref{sec:results-rq2}).
Table~\ref{tab:rq4-subjects} reports the subject-level IoU between each mutation approach and the real fault on the 12 subjects, and Fig.~\ref{fig:rq4} shows the distributions beneath these summaries.
TM-J achieves the highest descriptive median IoU (0.181), followed by DeepCrime (0.130), the norm-matched random control (0.117), and DeepMutation++ (0.058).
Two subjects, defect4ml068 and defect4ml069, yield zero IoU for every approach.
Both have degenerate test sets: the 200 test inputs of defect4ml068 contain only two distinct inputs, and defect4ml069 has only seven test inputs.

Table~\ref{tab:rq4-inference} reports the paired subject-level comparisons.
TM-J exceeds DeepMutation++ on nine of the ten subjects with a nonzero difference, with a median paired difference of 0.112.
This is the only comparison that remains significant after Holm correction ($p_{\mathrm{Holm}}=0.018$, rank-biserial $r=0.927$).
Against the norm-matched random control, TM-J also wins on nine of the ten non-tied subjects (median difference 0.067), but the difference is not significant after correction ($p_{\mathrm{Holm}}=0.168$).
The single loss against both comparators occurs on deepfd030, whose real fault causes only seven kills in total.
Against DeepCrime, TM-J wins on eight of the ten non-tied subjects, but the median paired difference is only 0.018 and the comparison is not significant ($p_{\mathrm{Holm}}=0.322$).
The positive descriptive margin over DeepCrime is concentrated in the defect4ML subjects.
Among the six subjects from CleanML, DeepFD, and DeepLocalize, TM-J is higher on four, by 0.003 to 0.045, whereas DeepCrime is higher on deepfd030 and deeplocalize030, the two with the fewest real-fault kills.

\begin{table}[t]
\centering
\caption{Real-fault similarity on the 12 subjects}
\label{tab:rq4-subjects}
\footnotesize
\setlength{\tabcolsep}{3pt}
\begin{tabular}{@{}lcccccc@{}}
\toprule
 & \multicolumn{4}{c}{IoU with real fault} & & DC \\
\cmidrule(lr){2-5}
Subject & TM-J & DC & Rand. & DM++ & KI & ceiling \\
\midrule
defect4ml038               & 0.248 & 0.008 & 0.117 & 0.010 & 486  & 0.008 \\
defect4ml061               & 0.172 & 0.105 & 0.116 & 0.131 & 120  & 0.131 \\
defect4ml067               & 0.257 & 0.175 & 0.135 & 0.057 & 7041 & 0.250 \\
defect4ml068               & 0.000 & 0.000 & 0.000 & 0.000 & 500  & 0.000 \\
defect4ml069$^\dagger$     & 0.000 & 0.000 & 0.000 & 0.000 & 12   & 0.000 \\
defect4ml100               & 0.202 & 0.132 & 0.122 & 0.065 & 2786 & 0.214 \\
cleanml018                 & 0.173 & 0.128 & 0.094 & 0.059 & 241  & 0.331 \\
cleanml019                 & 0.138 & 0.114 & 0.090 & 0.034 & 4702 & 0.225 \\
deepfd030$^\dagger$        & 0.143 & 0.333 & 0.347 & 0.219 & 7    & 0.707 \\
deepfd050                  & 0.195 & 0.184 & 0.115 & 0.086 & 2167 & 0.513 \\
deeplocalize030            & 0.189 & 0.308 & 0.154 & 0.041 & 85   & 0.515 \\
deeplocalize039            & 0.225 & 0.221 & 0.126 & 0.107 & 3028 & 0.478 \\
\midrule
Median                     & 0.181 & 0.130 & 0.117 & 0.058 &      &      \\
\bottomrule
\end{tabular}
\vspace{1mm}

\parbox{\columnwidth}{\footnotesize
DC: DeepCrime; Rand.: norm-matched random control; DM++: DeepMutation++.
KI: total kills of the real fault summed over its five instances.
DC ceiling: upper bound on DeepCrime's IoU set by the ratio of the smaller to the larger behavioral mass.
$^\dagger$Fewer than 50 real-fault kills.}
\end{table}

To examine why the approaches differ, we analyze the behavioral severity of their mutants on the ten subjects with at least 50 real-fault kills.
This floor keeps the analysis from resting on very sparse real-fault behavior: it excludes deepfd030 and defect4ml069, which have only seven and 12 kills, and any floor between 13 and 85 would exclude the same two subjects.
Since IoU is bounded by the ratio of the two behavioral masses (Section~\ref{sec:evaluation-measures}), Table~\ref{tab:rq4-subjects} also reports the resulting ceiling for DeepCrime.
Computed per configuration as IoU divided by the ceiling, DeepCrime reaches a median of 0.753 of its ceiling on the five defect4ML subjects; on defect4ml038, for example, its IoU and ceiling are both 0.008.
On the five subjects from CleanML, DeepFD, and DeepLocalize, it reaches only 0.479; on deepfd050, for example, it attains 0.184 against a ceiling of 0.513.
There, DeepCrime's similarity is limited more by which inputs kill its mutants than by their mass.
Relative to the real fault, TM-J carries more behavioral mass than DeepCrime on seven of the ten subjects and less on two, with one tie; the medians are 0.348 and 0.237, and the difference is not significant ($p=0.250$, rank-biserial $r=0.467$).
Both medians are below one, so the mutants of both approaches are typically milder than the real fault.

\begin{table}[t]
\centering
\caption{Paired subject-level comparisons of TM-J with each approach}
\label{tab:rq4-inference}
\footnotesize
\setlength{\tabcolsep}{4pt}
\begin{tabular}{@{}lccccc@{}}
\toprule
Comparator & W/L/T & Median $\Delta$ & $p$ & $p_{\mathrm{Holm}}$ & $r$ \\
\midrule
DM++      & 9/1/2 & 0.112 & 0.006 & 0.018 & 0.927 \\
Random    & 9/1/2 & 0.067 & 0.084 & 0.168 & 0.636 \\
DeepCrime & 8/2/2 & 0.018 & 0.322 & 0.322 & 0.382 \\
\bottomrule
\end{tabular}
\vspace{1mm}

\parbox{\columnwidth}{\footnotesize
W/L/T: wins, losses, and ties of TM-J across subjects. $\Delta$: IoU of TM-J minus IoU of the comparator, with the median taken over all 12 subjects.}
\end{table}

\begin{tcolorbox}[colback=gray!5, colframe=gray!60, boxrule=0.5pt, left=4pt, right=4pt, top=2pt, bottom=2pt]
\textbf{Answer to RQ4:}
TensorMutation (TM-J) achieves the highest median IoU with real faults (0.181), compared with 0.130 for DeepCrime, 0.117 for the norm-matched random control, and 0.058 for DeepMutation++.
Its advantage over DeepMutation++ is statistically significant after Holm correction.
Its paired differences from DeepCrime and the random control are positive but not statistically significant.
\end{tcolorbox}

\subsection{Does Source-Training Context Help?}
\label{sec:results-ablation}

Table~\ref{tab:ablation-context} compares TM-J with the two source-context variants defined in Section~\ref{sec:design-ablations}.
Neither variant improves on TM-J in any of the three measures.
On the test subjects, TM-J has the highest behavioral IoU and cosine similarity, followed by TM-J-Full and then TM-J-Narrow.
Against real faults, the order of the two variants reverses, but both remain below TM-J.
The two variants therefore rank differently on fidelity to DeepCrime and on similarity to real faults.
\begin{table}[t]
\centering
\caption{Source-context ablation}
\label{tab:ablation-context}
\footnotesize
\setlength{\tabcolsep}{5pt}
\begin{tabular}{@{}lccc@{}}
\toprule
 & \multicolumn{2}{c}{Test subjects} & Real faults \\
\cmidrule(lr){2-3}\cmidrule(lr){4-4}
Variant & IoU & Cosine & IoU \\
\midrule
TM-J        & 0.232 & 0.227 & 0.181 \\
TM-J-Full   & 0.223 & 0.214 & 0.140 \\
TM-J-Narrow & 0.170 & 0.195 & 0.155 \\
\bottomrule
\end{tabular}
\end{table}

\section{Discussion}
\label{sec:discussion}

\noindent Our key findings support one central conclusion: the behavioral effects of pre-training mutations can be learned as post-training weight-space transformations, and these transformations transfer to models never seen in training.
The resulting mutants show a higher median behavioral similarity to real faults than the pre-training reference, with no significant paired difference, and a significantly higher similarity than the post-training comparator.
Two limits come with this conclusion.
The learned direction adds less to the behavioral similarity than the medians suggest, once the magnitude that the model itself predicts is accounted for.
That magnitude is also not calibrated against the reference.
Below, we explain the reasons behind these results, discuss what they mean for realistic post-training mutation, and draw implications for testing and debugging of DL systems.

\subsection{Reasons for the Observed Results}
\label{sec:discussion-reasons}

We were initially surprised that TM-J reaches a higher median real-fault IoU than DeepCrime, the pre-training reference from which it learns.
The difference is not statistically significant, so we do not interpret this as TensorMutation surpassing its reference.
Still, the descriptive pattern requires interpretation, and the behavioral-severity analysis provides a partial explanation.
For two kill profiles, IoU is upper-bounded by the ratio of their smaller to larger behavioral mass.
For the five defect4ML subjects in the severity analysis, DeepCrime often operates close to this ceiling, indicating that limited behavioral mass constrains its attainable similarity.
By contrast, DeepCrime remains substantially below the ceiling on the subjects from CleanML, DeepFD, and DeepLocalize, so behavioral severity alone cannot explain the remaining gap.

DeepCrime's search strategy helps explain why its mutants are not arbitrarily severe.
DeepCrime searches for configurations that are killable but not trivial to kill, and for range-based parameters it uses binary search to identify such configurations~\cite{humbatova_deepcrime_2021}.
List-based parameters are instead searched exhaustively.
This design therefore avoids trivially strong mutants rather than maximizing behavioral severity.
Real faults are not calibrated to any such threshold.
On four of the five defect4ML subjects, the mass-imposed ceilings of 0.008 to 0.250 mean that the real faults carry roughly 4 to 120 times the behavioral mass of DeepCrime's mutants.
Why the benchmark faults are this severe, our data cannot say.
Our hypothesis is selection: faults with visible behavioral effects are more likely to be noticed and reported.

TensorMutation, in turn, tends to carry more behavioral mass than its DeepCrime reference.
On the nine held-out subjects, TM-J has greater behavioral mass than the corresponding DeepCrime mutants on seven subjects, with a median mass ratio of 1.366.
This is a behavioral calibration error against the reference.
Whether the additional mass helps against real faults depends on where it falls.
Mass on inputs that also kill the real fault raises the overlap, whereas mass on other inputs enlarges the union and lowers IoU, and even aligned mass stops helping once TM-J becomes more severe than the real fault.
Where TM-J remains less severe than the real fault, the additional mass raises the maximum IoU it can attain.
Consistent with this, TM-J carries more behavioral mass relative to the real fault than DeepCrime on seven of the ten subjects in the severity analysis, although the paired difference is not statistically significant.
Which inputs the mass falls on is a matter of direction.
RQ2 establishes that TensorMutation learns mutation-induced parameter direction, RQ3 shows that better directional fidelity tends to accompany better behavioral fidelity within subjects, and RQ4 provides supporting evidence against the norm-matched random control.

This account also implies that excess behavioral mass can become a disadvantage when the real fault itself is mild.
That pattern appears in the data.
Among the ten non-tied TM-J--DeepCrime comparisons, DeepCrime exceeds TM-J only on deepfd030 and deeplocalize030, the two subjects with the smallest real-fault kill counts.
We stress that this excess behavioral mass was found, not designed.
A direct follow-up would rescale DeepCrime's parameter deltas while preserving their direction and then re-evaluate behavioral mass and IoU, helping to separate the contributions of perturbation magnitude and direction.

The norm-matched random control isolates the contribution of directional information more directly.
It preserves TensorMutation's predicted perturbation magnitude while randomizing its direction, so its IoU captures the behavioral overlap obtainable at the same perturbation budget without the learned direction.
A systematic margin over this control is therefore evidence that the learned direction contributes information beyond perturbation magnitude.
On this basis we consider the direction claim established in parameter space.
In real-fault behavior it is supported but not statistically established.

\subsection{Learned Mutation and Mutant Realism} 
\label{sec:discussion-realism} 
 
Our earlier work on the competent programmer hypothesis found that even chains of mutations built from manually defined operators fail to reproduce most real faults, and it proposed learning mutations from real faults instead~\cite{ahmed_new_2024}. 
That proposal was made for conventional programs. 
TensorMutation explores this direction for DL by learning the effects of existing mutation operators derived from real-fault studies. 
The learned transformation does not add a new operator. 
It reproduces the effect of an existing one without the cost of retraining.
Our results show that these effects exhibit learnable structure that transfers to previously unseen models.

What the results support is narrower than what one might hope for.
Realism here means overlap of kill profiles with real faults.
This criterion rewards a match in severity.
If real-fault benchmarks favor faults with visible effects, as we hypothesized above, then mild mutants are penalized under it.
Mutants also serve a second role, as test goals, where resistant and hard-to-kill mutants are the valuable ones.
DeepCrime's search targets exactly such mutants.
Resistance and realism have been found to be largely independent qualities of a mutant~\cite{ahmed_quality-driven_2026}.
We make no claim about test-suite adequacy.
The two roles may therefore need differently calibrated generators.

The preceding paragraphs asked what the learned transformation produces.
The source-context ablation asks what it needs as input for better generalization.
Supplying the source model's training metadata did not improve either tested variant over TM-J.
One possible explanation is that the trained parameters already encode much of the outcome of training, so the metadata is largely redundant with the neural graph.
Another is that the FiLM pathway we tested does not transfer the supplied context to unseen subjects.
Our experiment cannot separate these explanations.
The defensible conclusion is narrower.
For the tested variants, conditioning on the trained graph and the requested mutation transferred better than also supplying the examined metadata.

Two of the three real-fault comparisons are not statistically significant, so what we conclude from RQ4 depends on how much weight they are given.
The medians consistently favor TensorMutation, but a non-significant paired difference establishes neither that TensorMutation exceeds DeepCrime nor that the two are equivalent.
We read the descriptive pattern as suggestive.
A more cautious reader would treat the comparison with DeepCrime as inconclusive.
Our evidence is consistent with that reading as well.
Neither reading changes the key result: TensorMutation is significantly more similar to real faults than the existing post-training mutation tool DeepMutation++.

\subsection{Implications}
\label{sec:implications}

For research on mutant realism, the main implication is methodological.
Studies of weight-space mutations that compare kill-profile overlap should include a norm-matched random control.
It keeps the predicted magnitude and randomizes only the direction, so it indicates how much of the overlap the learned direction adds.
In our case the answer is less than the medians suggest, which is exactly why the control belongs in the evaluation.
For the evaluation of weight-space methods, RQ3 shows that directional fidelity is associated with behavioral fidelity within subjects, but this relationship does not extend across subjects.
Behavioral evaluation therefore remains necessary, particularly when comparing results across subjects.

For testing and debugging research, TensorMutation generates mutants conditioned on fault-informed mutation specifications.
Each mutant retains its requested operator and configuration, allowing evaluations to distinguish, for example, learning-rate mutations from activation-function mutations.
Post-training perturbation tools also have identifiable operators, but these describe how the parameters are perturbed.
TensorMutation's operators correspond to developer-fault categories documented in empirical studies of real DL faults.
This provides a basis for studying fault-type-specific testing and debugging behavior.
However, RQ1 establishes only partial reproduction of the corresponding DeepCrime effects; fidelity to individual real-fault types requires further evaluation.

For practice, TensorMutation replaces per-mutant retraining with GNN inference and mutant reconstruction.
The trained GNN can be reused across supported models.
Although the statistical killing criterion still requires multiple instances per configuration, generating these mutants no longer requires retraining each one.

For the design of learned mutation, our results highlight the importance of calibration.
Although direct weight-space supervision improves directional fidelity, smaller parameter changes do not necessarily produce smaller behavioral effects.
Future approaches should therefore distinguish parameter magnitude from behavioral severity and consider how both affect similarity to the intended reference.
Weight-space learning also suggests applications beyond mutation.
Our corpus pairs original models with synthetic pre-training mutants, providing data that could be used to supervise fault-to-fix transformations.
Such transformations could support learned repair, while mutation-induced parameter changes may provide signatures for investigating fault localization.
Structural differences would require additional handling, and whether transformations learned from synthetic faults transfer to real faults remains untested.
We consider this a promising direction for future work.

\section{Threats to Validity}
\label{sec:threats-to-validity}
\noindent We discuss validity threats following the classification of Wohlin et al.~\cite{wohlin_experimentation_2024}.

\paragraph{Construct validity}
We measure mutant realism through behavioral similarity to real faults using kill-profile overlap.
This captures the dimension of realism that mutation testing generally evaluates.
Other downstream tasks may also require that a mutant and a real fault share the same root cause, which our measure does not capture.
Such correspondence is plausible, because the operators that supervise learning represent developer mistakes.
However, RQ1 and RQ2 show only partial fidelity to the corresponding DeepCrime mutants in behavior and in weight space.

\paragraph{Internal validity}
We identify three threats concerning the learning targets, the learning objective, and the data split.
First, the parameter deltas require position-by-position neuron correspondence between each original–mutant pair in the training set.
We generate each pair with matched per-instance seeds and deterministic GPU operations, so that a pair differs only by the mutation.
We verify each pair before building the targets, and identity mutants yield exactly zero deltas under this setup.
Second, the results depend on the learning objective.
We compare the behavioral objective alone with the joint objective under an identical recipe, but the coefficients were not tuned and each final variant was trained once.
Third, the same model can appear in more than one fault dataset.
We group subjects with identical architectures before splitting, keeping exact-model duplicates within the same partition.

\paragraph{Conclusion validity}
Our statistical analysis treats subjects rather than individual mutants as independent observations.
We use hierarchical aggregation and apply Holm correction to the three primary RQ4 comparisons.
However, the nine held-out subjects and 12 real-fault subjects limit statistical power, so non-significant differences cannot be interpreted as equivalence.

\paragraph{External validity}
Our evaluation covers MLP classifiers and 11 selected pre-training operators, each representing a fault type.
Generalization to different network families and other fault types remains untested.

\section{Conclusion}
\label{sec:conclusion}

\noindent This paper introduced TensorMutation, a learned post-training mutation approach that predicts the weight-space changes induced by pre-training mutations without retraining individual mutants.
Our evaluation shows that these parameter changes exhibit learnable structure that transfers to previously unseen models.
Direct weight-space supervision significantly improves directional fidelity, which is associated with behavioral fidelity within subjects but not across subjects.
In the real-fault evaluation, TensorMutation achieves a higher median behavioral similarity than DeepCrime, although the difference is not statistically significant.
It achieves significantly higher similarity than DeepMutation++, the evaluated post-training comparator.
Supplying the source model's training configuration as additional input did not improve the evaluated behavioral similarity or weight-space fidelity.

These findings demonstrate the potential of learning fault-informed weight-space transformations for post-training mutation.
Two limitations remain.
The norm-matched random control also achieves substantial real-fault similarity, so the additional behavioral contribution of the learned direction is not statistically established in the primary comparison.
Parameter magnitude and behavioral severity are also distinct quantities, and their calibration remains an open problem.
Future work will investigate controlled calibration, generalization to other network families, and the use of our paired corpus to learn fault-to-fix transformations for repair and fault localization.
We also plan to develop TensorMutation into a reusable mutant-generation tool for research on DL testing and debugging.

\section*{Data and Code Availability}
\noindent The visualization notebook and the result CSV files underlying the reported analyses are available at \url{https://doi.org/10.5281/zenodo.23318504}.
The full replication package will be made publicly available upon acceptance.

\section*{Acknowledgment} 
 
\noindent The authors used Claude and ChatGPT during the preparation of this manuscript to assist with language refinement, clarity, presentation of the content, and software implementation tasks. 
All scientific ideas, decisions, experimental designs, analyses, interpretations, and conclusions remain the work of the authors, who take full responsibility for the content.

%{\appendix[Proof of the Zonklar Equations]
%Appendix here.
%}



\bibliographystyle{IEEEtran}
\bibliography{references}

\end{document}